\documentclass{webofc}

\usepackage[varg]{txfonts}   
\usepackage{hyperref}
\usepackage{url}

\usepackage{graphicx}
\usepackage{subcaption}
\usepackage{wrapfig}
\hypersetup{colorlinks=true,citecolor=blue,urlcolor=blue,linkcolor=blue}
\begin{document}
\title{Towards an Introspective Dynamic Model of Globally \\Distributed Computing Infrastructures}
%
%

\author{
    \firstname{Ozgur O.} \lastname{Kilic}\inst{1} \and
    \firstname{David K.} \lastname{Park}\inst{1} \and
    \firstname{Yihui} \lastname{Ren}\inst{1} \and
    \firstname{Tatiana} \lastname{Korchuganova}\inst{3} \and
    \firstname{Sairam Sri} \lastname{Vatsavai}\inst{1} \and
    \firstname{Joseph} \lastname{Boudreau}\inst{3} \and
    \firstname{Tasnuva} \lastname{Chowdhury}\inst{1} \and
    \firstname{Shengyu} \lastname{Feng}\inst{4} \and
    \firstname{Raees} \lastname{Khan}\inst{3} \and
    \firstname{Jaehyung} \lastname{Kim}\inst{4} \and
    \firstname{Scott} \lastname{Klasky}\inst{2} \and
    \firstname{Tadashi} \lastname{Maeno}\inst{1} \and
    \firstname{Paul} \lastname{Nilsson}\inst{1} \and
    \firstname{Verena Ingrid Martinez} \lastname{Outschoorn}\inst{5} \and
    \firstname{Norbert} \lastname{Podhorszki}\inst{2} \and
    \firstname{Fr\'ed\'eric} \lastname{Suter}\inst{2} \and
    \firstname{Wei} \lastname{Yang}\inst{6} \and
    \firstname{Yiming} \lastname{Yang}\inst{4} \and
    \firstname{Shinjae} \lastname{Yoo}\inst{1} \and
    \firstname{Alexei} \lastname{Klimentov}\inst{1} \and
    \firstname{Adolfy} \lastname{Hoisie}\inst{1}
}

\institute{
    Brookhaven National Laboratory, Upton, NY, USA  \and
    Oak Ridge National Laboratory, Oak Ridge, TN, USA \and
    University of Pittsburgh, Pittsburgh, PA, USA \and
    Carnegie Mellon University, Pittsburgh, PA, USA \and
    University of Massachusetts, Amherst, MA, USA \and
    SLAC National Accelerator Laboratory, Menlo Park, CA, USA
}

\abstract{Large-scale scientific collaborations like ATLAS, Belle II, CMS, DUNE, and others involve hundreds of research institutes and thousands of researchers spread across the globe. These experiments generate petabytes of data, with volumes soon expected to reach exabytes. Consequently, there is a growing need for computation, including structured data processing from raw data to consumer-ready derived data, extensive Monte Carlo simulation campaigns, and a wide range of end-user analysis. To manage these computational and storage demands, centralized workflow and data management systems are implemented. However, decisions regarding data placement and payload allocation are often made disjointly and via heuristic means. A significant obstacle in adopting more effective heuristic or AI-driven solutions is the absence of a quick and reliable introspective dynamic model to evaluate and refine alternative approaches.
In this study, we aim to develop such an interactive system using real-world data. By examining job execution records from the PanDA workflow management system, we have pinpointed key performance indicators such as queuing time, error rate, and the extent of remote data access. The dataset includes five months of activity. Additionally, we are creating a generative AI model to simulate time series of payloads, which incorporate visible features like category, event count, and submitting group, as well as hidden features like the total computational load—derived from existing PanDA records and computing site capabilities. These hidden features, which are not visible to job allocators, whether heuristic or AI-driven, influence factors such as queuing times and data movement.
\vspace{-1cm}
}
\maketitle
\section{Introduction}
\label{sec:intro}

\begin{figure*}[b]  
    \centering
    \includegraphics[width=0.7\linewidth]{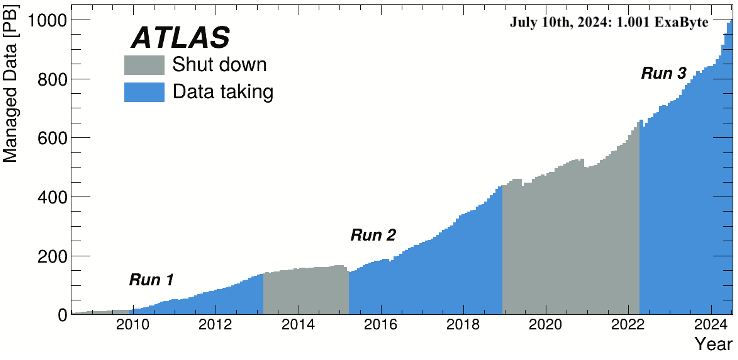}
    \vspace*{-0.4cm} 
    \caption{Total amount of data generated by ATLAS experiment reported by RUCIO~\cite{rucio}}
    \label{fig:atlas_data}       
\end{figure*}

Large-scale scientific collaborations in high-energy physics (HEP), such as ATLAS~\cite{atlas,atlas_data_processing_chain,barreiro2017atlas}, CMS~\cite{cms2008cms}, and DUNE~\cite{falcone2022deep}, are generating unprecedented volumes of data (e.g., exceeding exabyte for ATLAS, see Figure~\ref{fig:atlas_data}) which demand innovative approaches to data processing and resource management. In this context, the detailed analysis of operational data becomes essential for identifying inefficiencies and unlocking new avenues for optimization.

Our work focuses on an exceptionally large dataset collected from the PanDA workflow management system~\cite{panda}, spanning five months of continuous operations and comprising millions of job execution records. This dataset captures a rich set of performance metrics including job queuing times, error rates, remote data accesses, and more. The magnitude and granularity of this dataset do not only provide a robust foundation for empirical analysis but also reveal latent patterns that traditional heuristic approaches often overlook. However, the conventional reliance on heuristic methods for data placement and payload allocation often fails to capture the dynamic interactions inherent in such large-scale systems.

The advantages of our comprehensive analysis are multifold. By quantifying key performance indicators, we can pinpoint specific bottlenecks in resource allocation and data placement, leading to improved throughput and system resilience. This data-driven approach enables more informed decision-making and prioritizes system improvements based on empirical evidence, potentially reducing idle times and optimizing overall performance.

Complementing our analysis, we introduce  how to use generative AI models~\cite{scAIsurrogate2025} that are capable of synthesizing realistic time series tabular data that mirrors both the observable characteristics and hidden dynamics like the inferred total computational load. 
This model not only reproduces the complex statistical distribution of the observed data but could also be used to facilitate the exploration of “what-if” scenario simulations. 
Such simulations would provide a risk-free environment to evaluate alternative resource allocation strategies and predict the impact of potential changes.

By integrating an in-depth data analysis with advanced generative modeling, our study bridges the gap between static, heuristic decision-making and dynamic, data-driven optimization. The insights derived from this dual approach are expected to significantly contribute to the enhancement of efficiency, reliability, and sustainability in distributed high-performance computing systems for large-scale scientific collaborations.

\section{Related Work}
\label{sec:related}

The literature on workflow management and dynamic system modeling is extensive. High-profile studies from large-scale collaborations~\cite{atlas_computing,taylor2015evolution,karpenko2015atlas} have characterized the operational challenges of distributed computing environments, focusing on job queuing delays, error propagation, and resource underutilization. For instance, early works on PanDA~\cite{atlas_computing,panda} provided foundational insights into distributed workload management and served as a stepping stone for subsequent research into system performance metrics. We collected a large dataset from PanDA about ATLAS jobs, tasks, and data movements for a five-month period and did various analyses using that dataset, which has more extensive information.

Tabular data consist of structured tables with both categorical and numerical features but are often limited in size compared to vision or natural language processing (NLP) datasets. The growing demand for high-quality synthetic tabular data has driven research into deep generative models, leading to a competitive performance in data augmentation~\cite{xu2019modeling, kotelnikov2023tabddpm}. Popular approaches include autoencoders, generative adversarial networks (GANs), transformers, and diffusion models, alongside heuristic methods like the Synthetic Minority Over-sampling Technique (SMOTE). Despite its simplicity, SMOTE has been reported to be highly efficient and effective in capturing tabular distributions, making it a widely used technique for generating synthetic tabular data~\cite{chawla2002smote}.

\begin{figure*}[ht]
\centering
\includegraphics[width=0.95\linewidth]{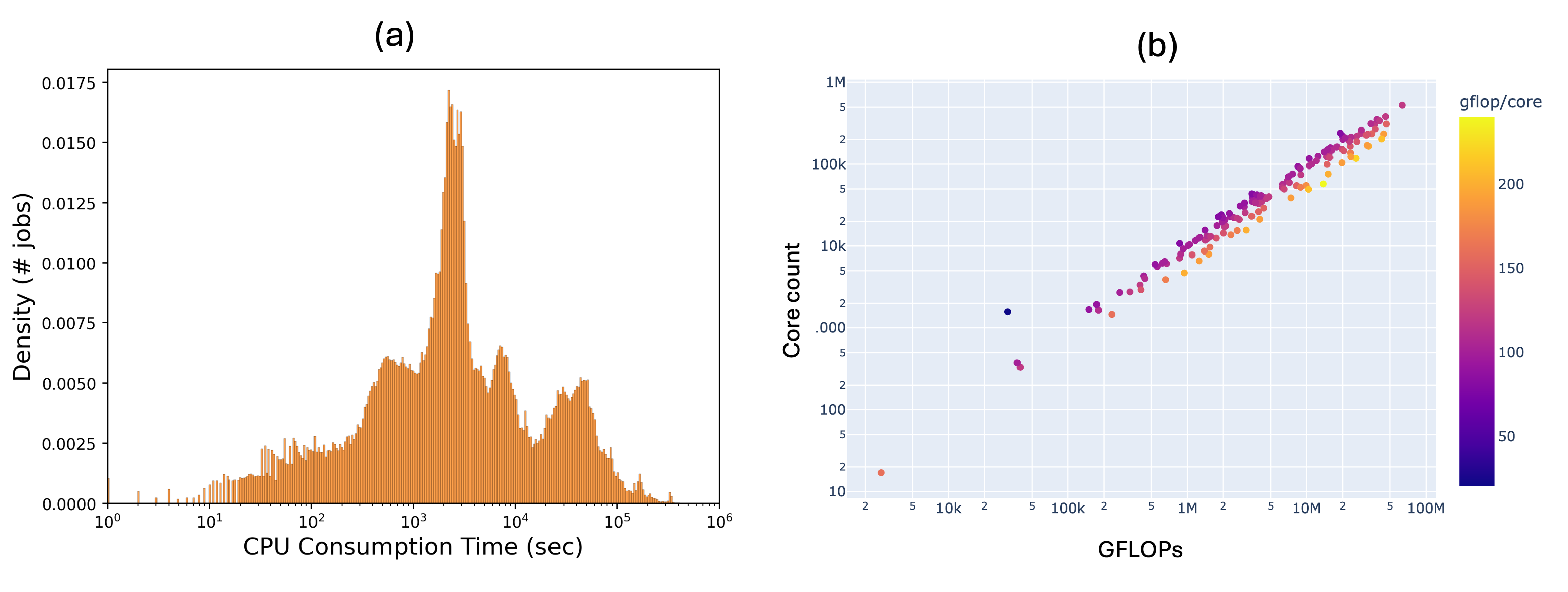}
\vspace*{-0.3cm}
\caption{(a) Distribution of CPU Consumption across jobs, and (b) GFLOP vs. Core Count across all sites}
\label{fig:comparison} 
\vspace{-0.8cm}
\end{figure*}


\begin{figure}[ht]
    \centering
    \begin{subfigure}{0.45\linewidth}
        \centering
        \includegraphics[width=\linewidth]{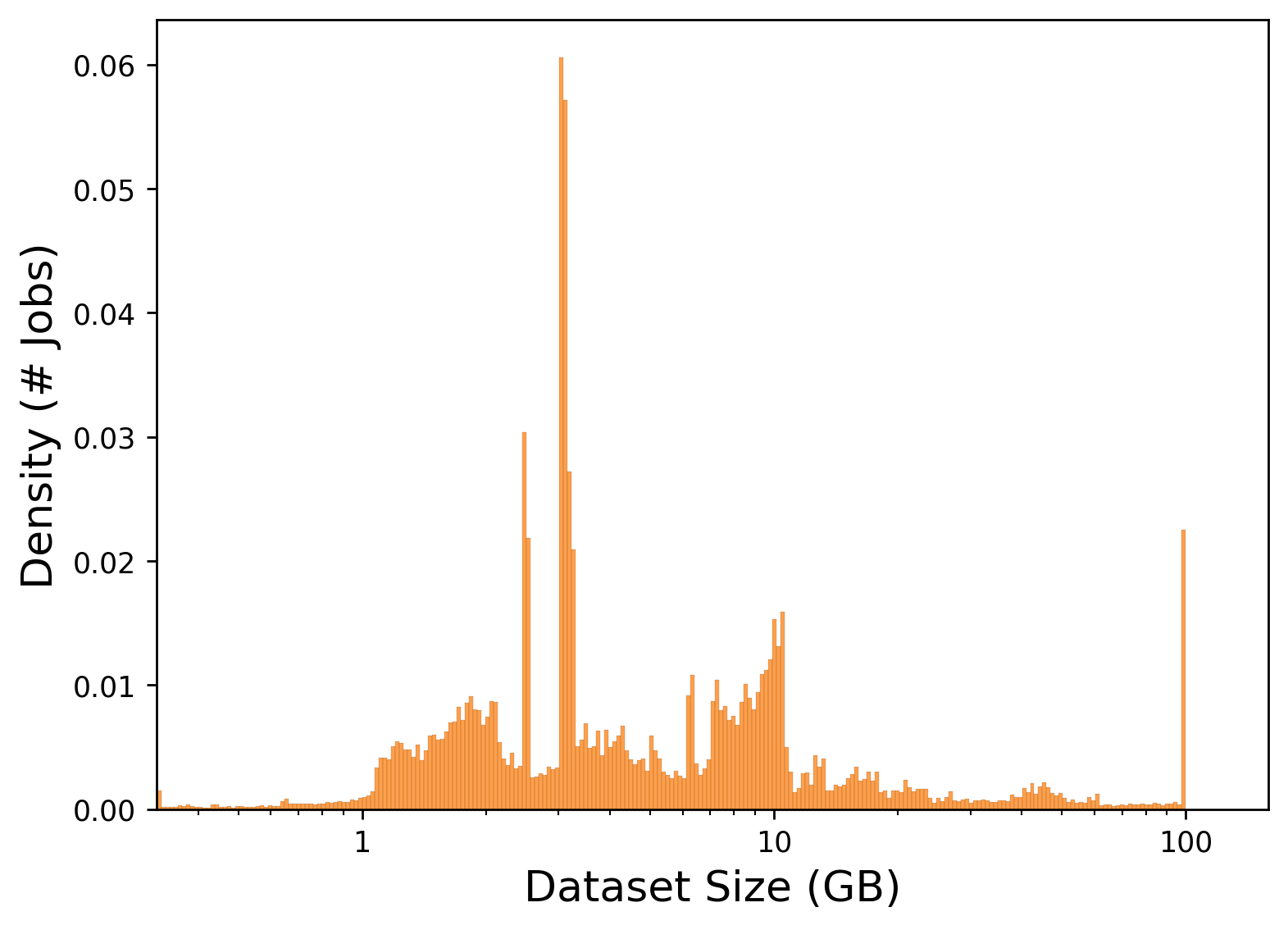}
        \caption{Dataset Size distribution}
        \label{fig:dataset_size}
    \end{subfigure}
    \hfill
    \begin{subfigure}{0.45\linewidth}
        \centering
        \includegraphics[width=0.85\linewidth]{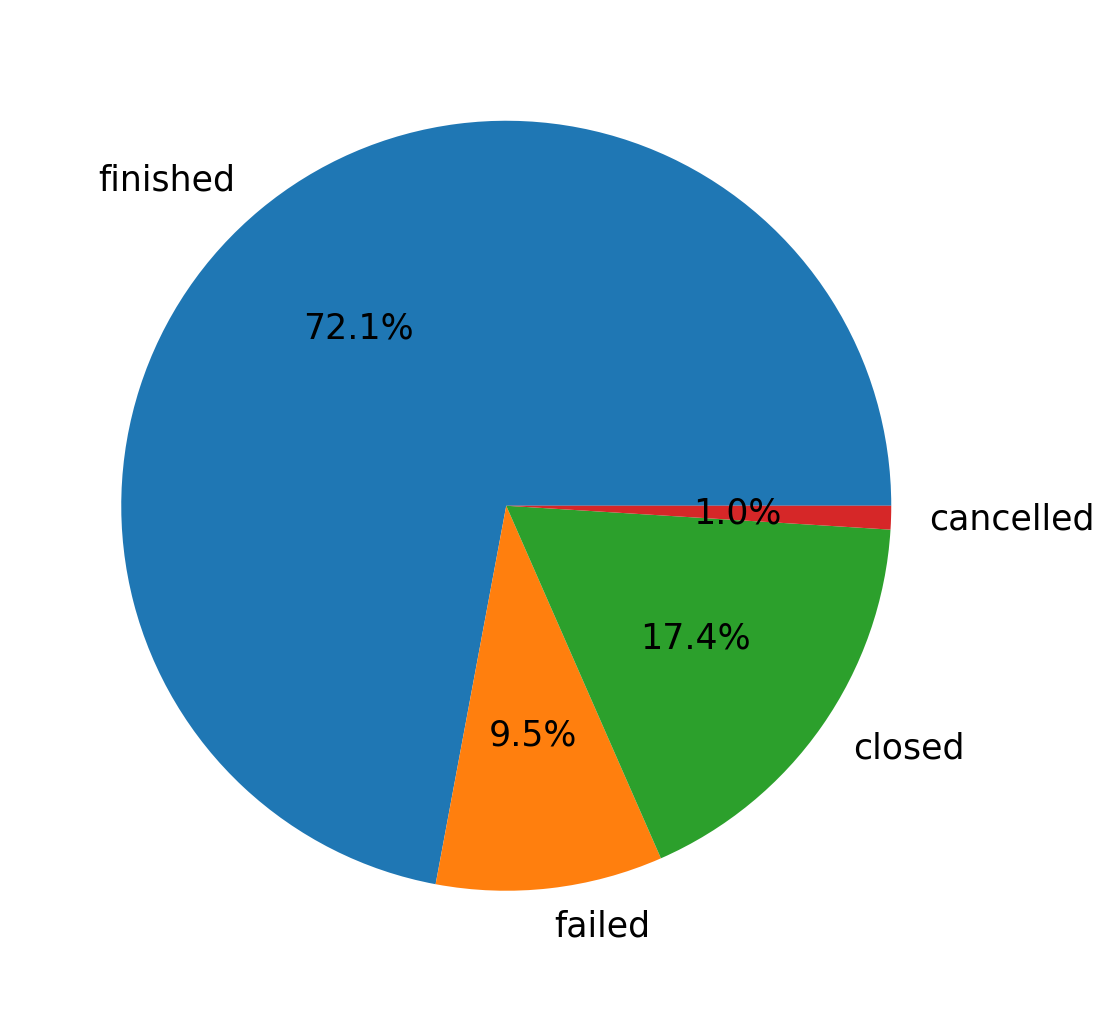}
        \caption{Distribution of Job Status}
        \label{fig:job_errors_pie}
    \end{subfigure}
    \caption{Analysis of Data (a) and Job status(b) Distribution}
    \label{fig:data_job_distribution}
    \vspace*{-0.8cm}
\end{figure}


\begin{figure}[ht]
    \centering
    \begin{subfigure}{0.45\linewidth}
        \centering
        \includegraphics[width=\linewidth]{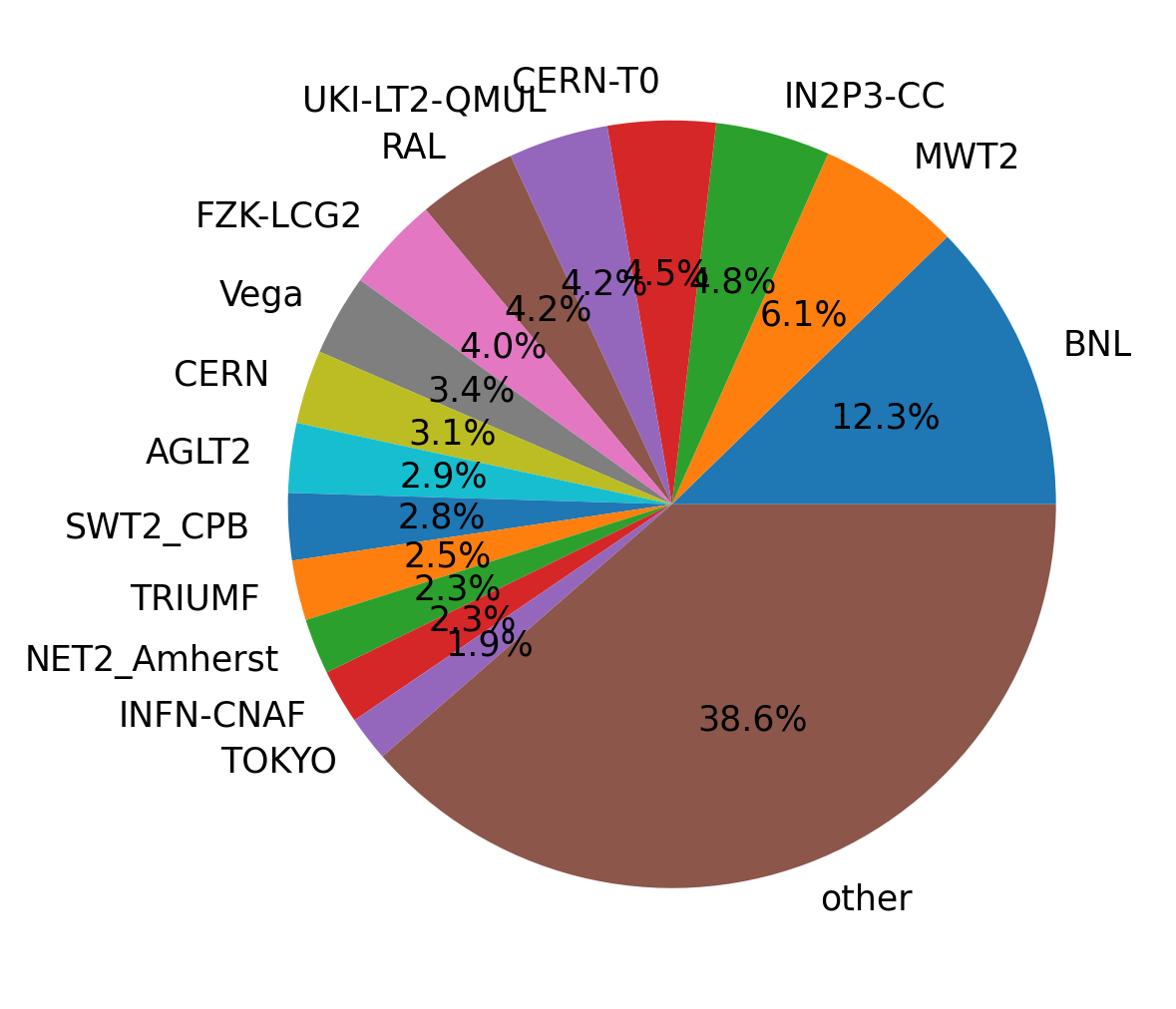}
        \caption{Job Count by Site}
        \label{fig:job_count_pie_by_site}
    \end{subfigure}
    \hfill
    \begin{subfigure}{0.45\linewidth}
        \centering
        \includegraphics[width=\linewidth]{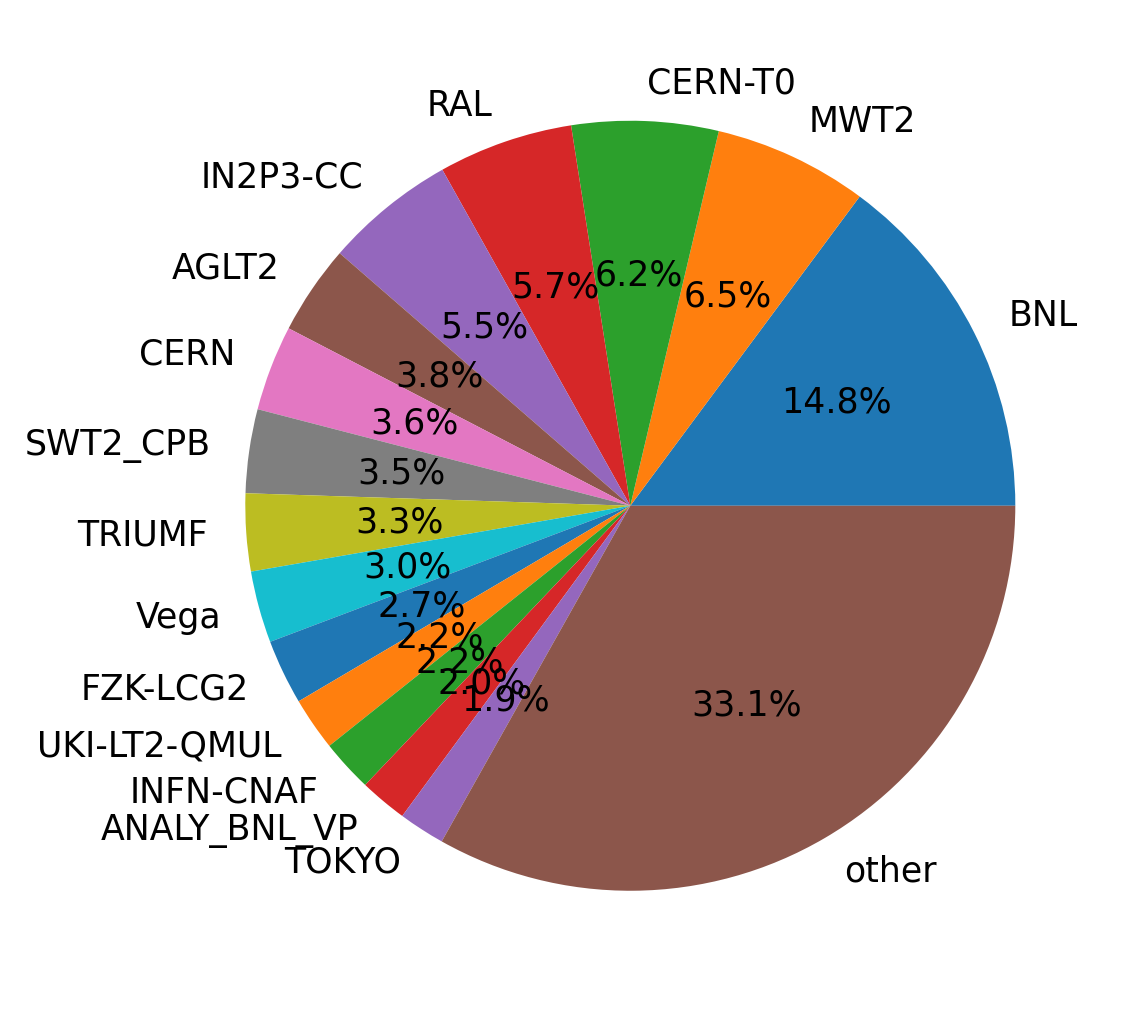}
        \caption{Dataset Size by Site}
        \label{fig:dataset_size_pie_by_site}
    \end{subfigure}
    \caption{Distribution of Job Count and Dataset Size by Site}
    \label{fig:site_distribution}
    \vspace*{-0.8cm}
\end{figure}

\section{PanDA records}
\label{sec:data}
The foundation of our study is the extensive dataset collected from the PanDA workflow management system. This dataset, spanning a continuous period of five months, comprises task and job execution records, dataset locations and usage, along with multiple key performance indicators: (i)
Job Queuing Time -- the delay between job submission and execution start; (ii) Various Error Reasons and Results -- differentiated by error types such as pilot and dispatcher errors and how they affected the execution of the job or tasks; (iii) Remote Data Access -- Data access locations; and (iv) Resource Utilization Metrics -- Total execution time, queue time, number of CPUs, and I/O.

Data was acquired directly from the operational logs of PanDA. After collection, we processed the data to categorize visible features (e.g., job type, submitting group) and compute hidden features (e.g., total computational load estimated from site capabilities). The resulting dataset offers a rich, multifaceted view of the operational dynamics, suitable for both descriptive analysis and training generative models. Data analysis is discussed in Section~\ref{sec:analysis} and data generation using generative AI is discussed in Section~\ref{sec:genAI}.

\section{Data Analysis}
\label{sec:analysis}

This section provides an in-depth examination of the five-month dataset collected from the PanDA workflow management system. We focus on four main areas: (i) overall data volume and context, (ii) job execution characteristics, (iii) error analysis and job status, and (iv) job queue times and wasted resource analysis. By dissecting these aspects, we aim to identify performance bottlenecks and inefficiencies in globally distributed computing environments. This analysis will also guide the development of an introspective model for optimizing resource allocation and data placement.


\subsection{Data Volume and Context}
As shown in Figure~\ref{fig:atlas_data} the total amount of data generated by the ATLAS experiment continues to grow rapidly and is projected to reach exabyte scales~\cite{rucio}. This figure underscores the sheer scale of HEP data, which is on track to reach exabyte levels in upcoming years. Such magnitude necessitates robust data-management strategies and motivates our effort. For this reason we collected five months of execution and data movement data of ATLAS experiment from PanDA and Rucio. The collected data consist of various different aspects of a task/job execution cycle. For this study we focused on three categories: Job Execution (status, execution time, queue time, CPU consumption, Error, etc), Tasks Execution (similar metrics as jobs), and Datasets (location, size, etc).


\subsection{Job Execution and Grid}
One of the first goals of our analysis was to understand how job runtimes and site capabilities affect overall performance. Figure~\ref{fig:comparison}a illustrates a high-level distribution of CPU consumption for a large sample of jobs. Most jobs consume on the order of thousands of CPU-seconds, reflecting the computationally intensive nature of high-energy physics workflows. Meanwhile, Figure~\ref{fig:comparison}b plots the relationship between the core count of a site and its theoretical GFLOPs, showing that sites with higher core counts also provide correspondingly higher total GFLOPs. Only a small subset of sites exhibits notably high GFLOPs per core (highlighted as yellow dots).

To provide further insight into data volumes, Figure~\ref{fig:dataset_size} depicts the overall size distribution of datasets. The data reveal that while many datasets remain relatively modest in size (on the order of gigabytes), a nontrivial fraction of them reach tens or even hundreds of gigabytes. These large datasets often drive both higher queue time and increased network traffic, especially if they must be transferred from remote sites for processing.

Beyond raw CPU consumption and dataset sizes, we also investigated how workloads and data are distributed across the ATLAS grid. Figure~\ref{fig:job_count_pie_by_site} illustrates the number of jobs executed at each site during the five-month period, while Figure~\ref{fig:dataset_size_pie_by_site} highlights the total dataset size handled by each site. Some sites, typically Tier-1 centers (e.g., BNL), show a markedly higher volume of jobs and larger dataset footprints. Understanding this distribution is critical for modeling load-balancing strategies and predicting bottlenecks in job dispatch. Note that there are more than 170 sites, and we combined all the sites that have lower than 1.9\% as ``Others''. 


\begin{figure*}[t]  
    \centering
x    \includegraphics[width=\linewidth]{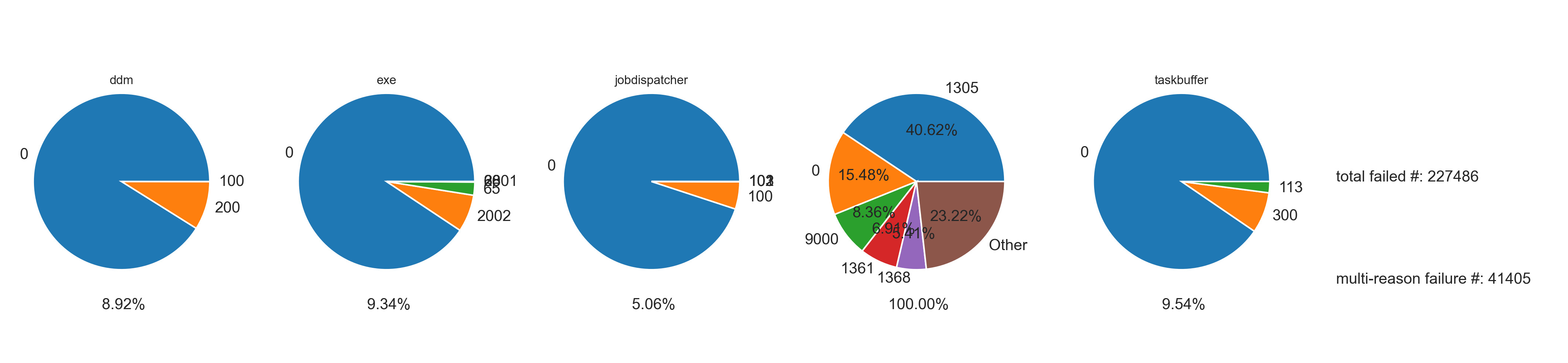}
    \vspace*{-0.8cm} 
    \caption{Job Error Distribution for PanDA record. This study done using recorded 5 months metadata and the numbers shown on pie chart are the select error codes. We also show total number of jobs failed  and multi reason failures at the end of the figure }
    \label{fig:job_errors_pie}
    \vspace*{-0.8cm} 
\end{figure*}

\subsection{Error and Status Analysis}
Effective resource utilization hinges not only on raw compute capacity but also on minimizing job failures and mitigating systemic failures. Figure~\ref{fig:job_errors_pie} shows the distribution of the different error categories observed in our dataset. These errors include pilot-related issues, file transfer failures, and job dispatcher errors, among others. Pilot errors, for instance, often arise from running code, software or hardware environment problems on worker nodes, memory issues, while dispatcher errors typically occur before jobs are allocated to a given pilot. Note that the total percent of errors is  $\approx117\%$ because a job may fail due to multiple different errors which will be marked with multiple error codes. 

Complementing this error breakdown, Figure~\ref{fig:job_count_pie_by_site} details the overall distribution of job statuses -  finished (successfully finished), failed (gave an error), closed, and canceled (terminated either by user of system).  Although a majority of jobs are successfully completed (finished, closed or canceled), failed jobs still constitute a significant fraction of the total workload. As a future work, by correlating failure modes with site characteristics and job parameters, we can pinpoint recurring patterns—such as consistently high failure rates at specific sites or error codes disproportionately triggered by large datasets or errors caused by specific user/group. These insights feed directly into our ongoing modeling efforts, where we aim to predict and reduce failure rates via improved data placement and scheduling policies.

\subsection{Job Queues and Wasted Time Analysis}
\begin{figure}[ht]
    \centering
    \begin{subfigure}{0.49\linewidth}
        \centering
        \includegraphics[width=\linewidth]{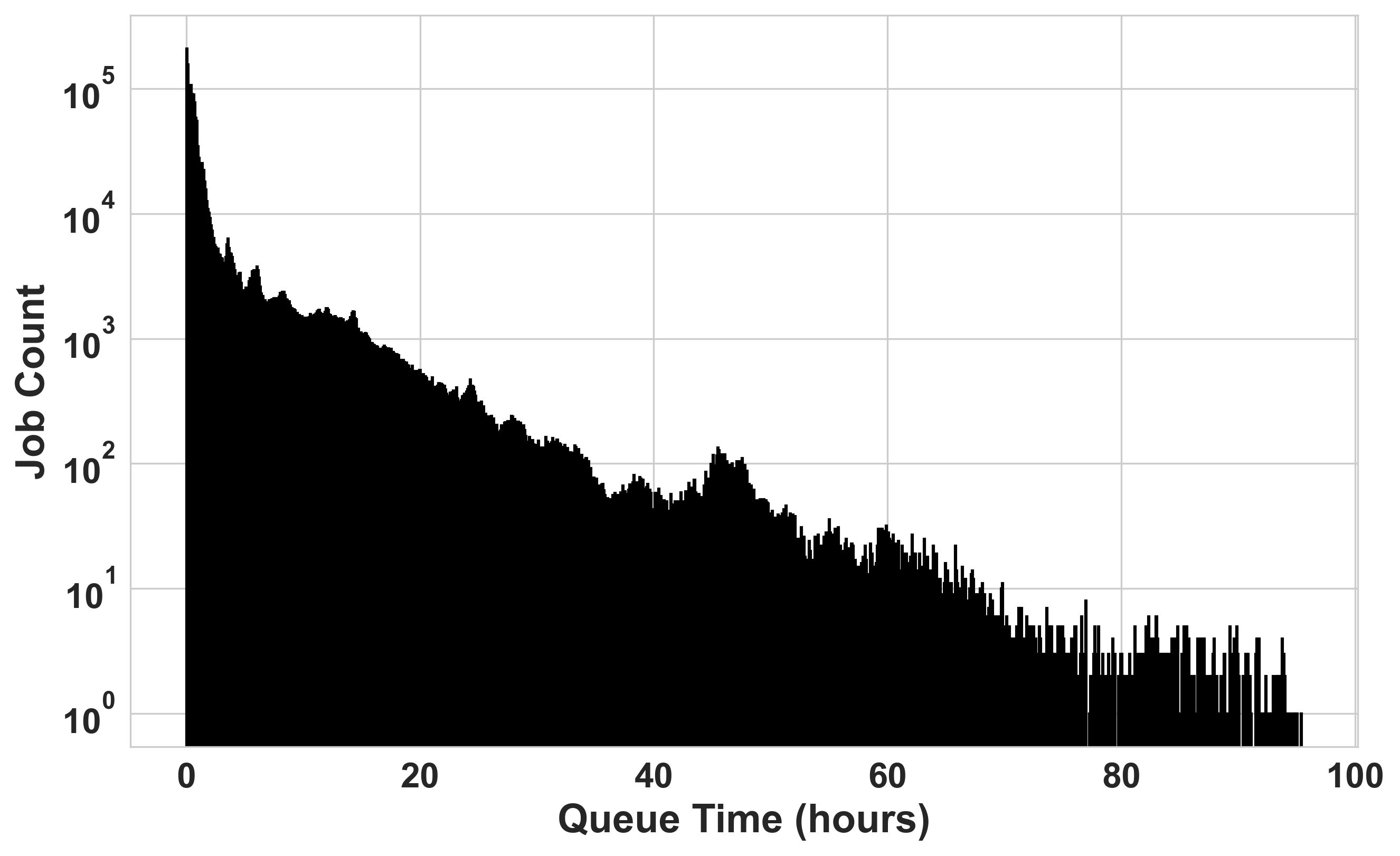}
        \caption{User Jobs Queuing Time Histogram}
        \label{fig:queue_time_hist}
    \end{subfigure}
    \hfill
    \begin{subfigure}{0.47\linewidth}
        \centering
        \includegraphics[width=\linewidth]{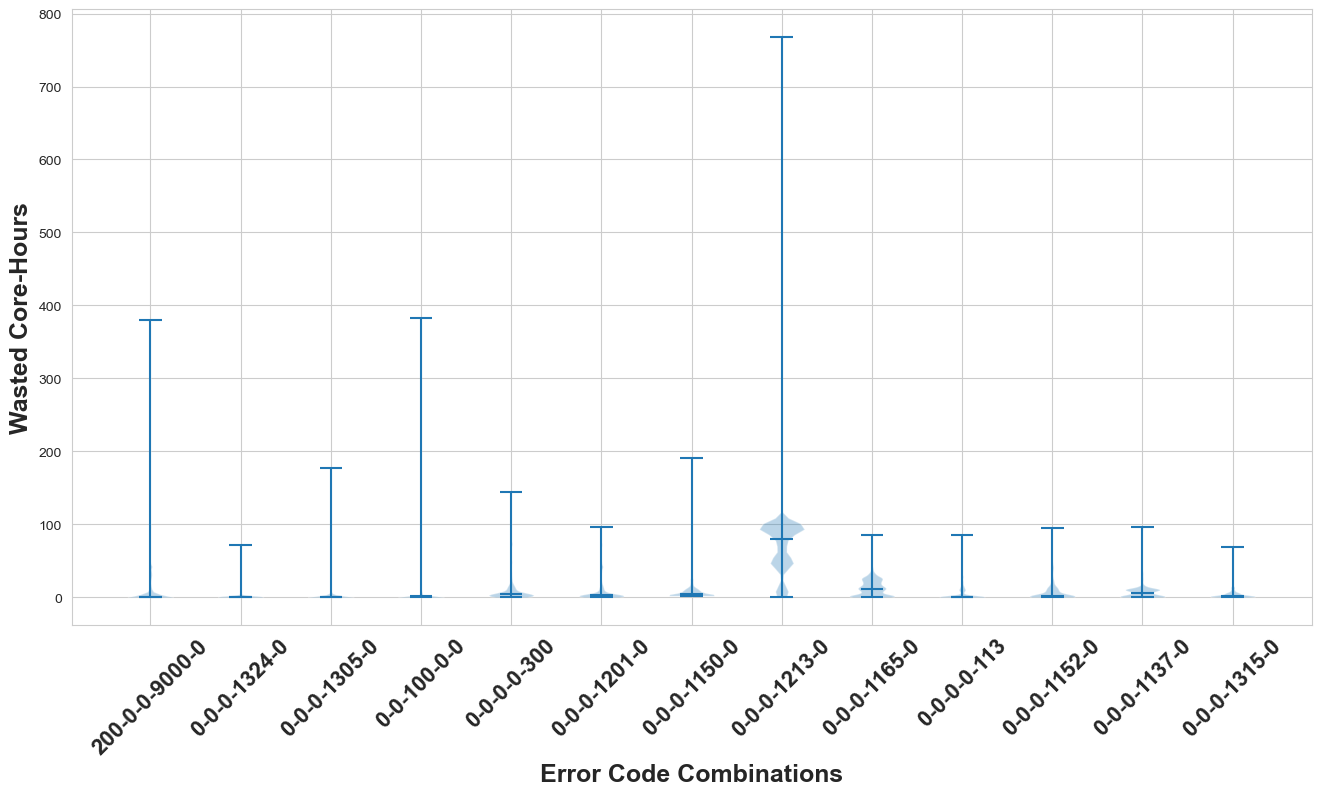}
        \caption{Wasted Core Hours for Error Combination}
        \label{fig:wasted_core_hours_violin}
    \end{subfigure}
    \caption{Analysis of User Jobs: Queuing Time and Wasted Core Hours}
    \label{fig:user_jobs_analysis}
    \vspace*{-0.8cm}
\end{figure}

Long queuing times and wasted CPU cycles further constrain the efficiency of distributed systems. Figure~\ref{fig:queue_time_hist} focuses on user jobs, plotting a histogram of their queuing times from submission to the start of execution. While many user jobs begin running within a relatively short window (within a day or two), there is a noticeable tail of jobs with much longer waits, potentially due to site unavailability or data not being co-located. Such delays can cascade, causing idle resources at other sites or user workflows to stall. This could possibly be improved by a more knowledgeable data and job allocation paradigm.



An equally critical metric is the amount of CPU time wasted by failed jobs. Figure~\ref{fig:wasted_core_hours_violin} visualizes wasted core-hours as a function of different error-code combinations. For example, PAYLOADEXECUTIONFAILURE error (1305 4th bar) and REMOTEFILECOULDNOTBEOPENED error (1361 19th bar)~\cite{panda_wms_errorcodes}  can accumulate many hours of runtime before the failure is triggered. Identifying these problematic error codes enables targeted interventions to reduce wasted cycles.

Overall, the patterns observed in queuing times and wasted CPU hours emphasize the importance of a more dynamic and introspective approach to job dispatch and data management. By addressing site-level mismatches, proactively replicating data to reduce remote transfers, and focusing on high-impact error codes, we can significantly improve the throughput and reliability of the system.


\begin{figure*}
\centering
\includegraphics[width=0.9\linewidth]{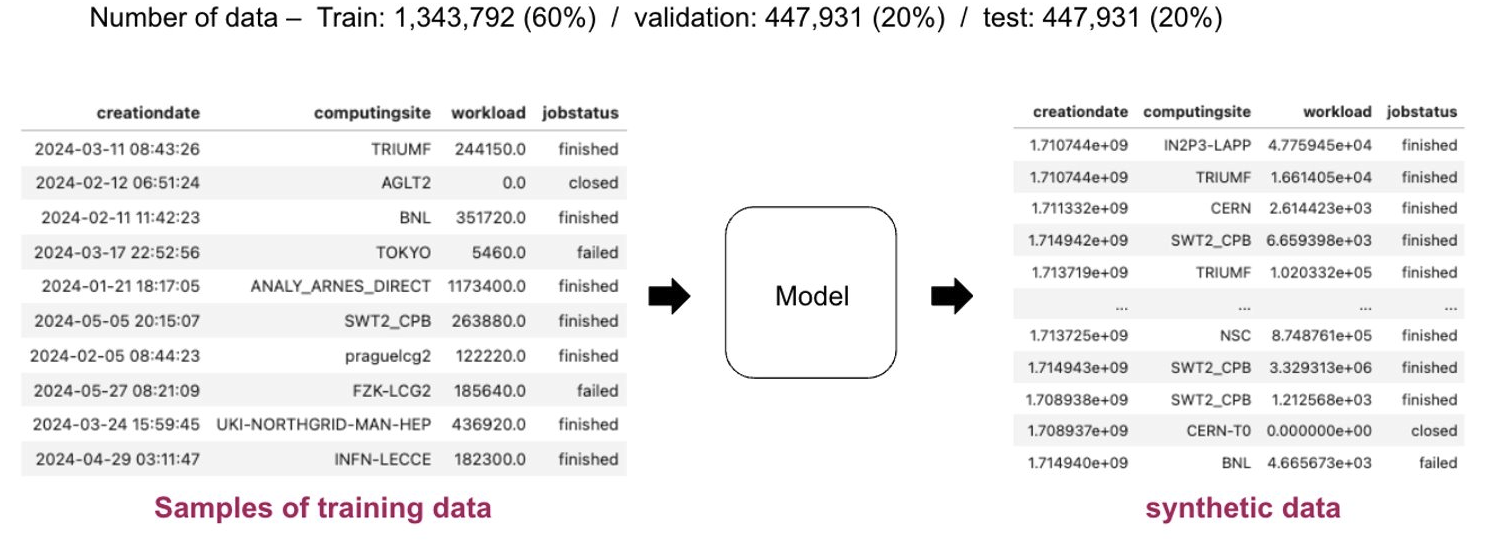}
\caption{Dataset statistics and generative modeling scheme using PanDA~\cite{panda} records in tabular data format. Our goal is to build a surrogate model for PanDA data, which is capable to represent the data distribution and synthesize new data. This is a sampled view: we use nine variables in total for the evaluation. }
\label{fig:generativeAI} 
\vspace*{-0.8cm} 
\end{figure*}

\section{Data Generation with Generative AI}
\label{sec:genAI}

PanDA records are structured as tabular data, comprising both categorical and numerical variables. In total, we downsample nine variables: \emph{creationtime}, which represents the timestamp when a job is created, and \emph{computingsite}, which indicates where the job is executed. Additionally, five dataset-related attributes are included: \emph{project} (project name), \emph{prodstep} (production step), \emph{datatype} (dataset type), \emph{ninputdatafiles} (number of input files, also referred to as \emph{nfiles}), and \emph{inputfilebytes} (gross input size, also referred to as \emph{size}). These seven features are known before execution, whereas the last two variables, \emph{jobstatus} and \emph{workload} are determined only after the job is completed, with \emph{workload} computed as the product of the number of cores, Gflop per core, and CPU time used. Given the complex dependencies among these variables, generating high-fidelity synthetic tabular data requires capturing intricate relationships between categorical and numerical features.

\begin{figure*}[h!]
\centering
\includegraphics[width=0.9\linewidth]{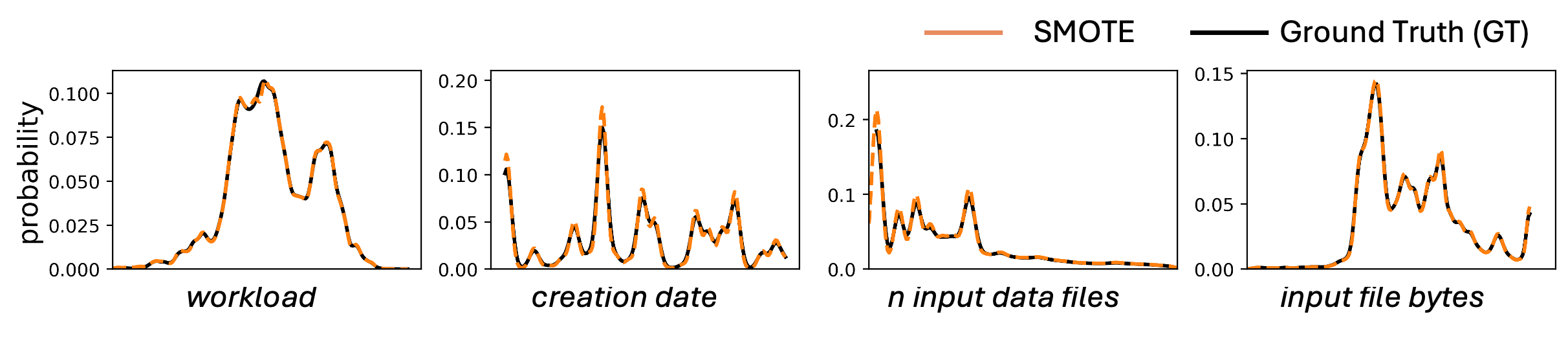}
\vspace*{-0.2cm} 
\caption{Generative performance of SMOTE. Distributions of four numerical columns are compared against the ground truth training data}
\label{fig:genai}       
\vspace*{-0.5cm} 
\end{figure*}

To tackle this challenge, we use SMOTE, an interpolation-based method commonly applied to generate synthetic samples in imbalanced datasets. SMOTE is trained by selecting a sample from the minority class, identifying its nearest neighbors based on a chosen distance metric (e.g., Euclidean distance for numerical features or Hamming distance for categorical ones), and generating synthetic data points along the vector connecting the original data to its neighbors. This is achieved by randomly interpolating feature values while preserving the underlying structure of the dataset. For categorical features, SMOTE assigns the most frequent neighbor category rather than interpolating. Once trained, SMOTE synthesizes data matching the size of the original dataset while preserving its statistical properties. To evaluate generative quality, we compare the distribution of individual features in the synthetic dataset to the ground truth. Results in Figure~\ref{fig:genai} indicate that SMOTE effectively learns the distribution of each feature, demonstrating its capability to generate realistic synthetic PanDA records.

\section{Conclusion}
\label{sec:conc}
In this paper, we presented a comprehensive data-driven analysis of large-scale distributed computing in the context of high-energy physics workflows. By examining five months of PanDA logs, we identified key performance bottlenecks—such as high failure rates, long queuing times, and significant remote data transfers—that collectively impact overall system efficiency. We also introduced an approach for generating synthetic workload data using AI-driven techniques, which enables “what-if” scenario evaluations without risking production stability. By integrating real-world observations with generative models, our work lays the groundwork for more adaptive and introspective resource allocation and data management strategies. Future research will focus on refining these surrogate models, extending our analysis to additional sites and data types, and ultimately improving the reliability and sustainability of globally distributed computing infrastructures.





\bibliography{ref}

\end{document}